\documentclass[12pt,a4paper]{article}
\makeatletter
\def\eqnarray{%
\stepcounter{equation}%
\let\@currentlabel=\theequation
\global\@eqnswtrue
\global\@eqcnt\z@
\tabskip\@centering
\let\\=\@eqncr
$$\halign to \displaywidth\bgroup\@eqnsel\hskip\@centering
$\displaystyle\tabskip\z@{##}$&\global\@eqcnt\@ne
\hfil$\displaystyle{{}##{}}$\hfil
&\global\@eqcnt\tw@$\displaystyle\tabskip\z@{##}$\hfil
\tabskip\@centering&\llap{##}\tabskip\z@\cr}
\makeatother
\newcommand{\fukuso}{{\mathbf C}}

\begin{document}

\title{\sl Explicit Form of the Evolution Operator for the Four Atoms 
Tavis--Cummings Model}
\author{
  Kazuyuki FUJII
  \thanks{E-mail address : fujii@yokohama-cu.ac.jp },\quad 
  Kyoko HIGASHIDA 
  \thanks{E-mail address : s035577d@yokohama-cu.ac.jp },\quad 
  Ryosuke KATO 
  \thanks{E-mail address : s035559g@yokohama-cu.ac.jp }\\
  Tatsuo SUZUKI
  \thanks{E-mail address : suzukita@gm.math.waseda.ac.jp },\quad 
  Yukako WADA 
  \thanks{E-mail address : s035588a@yokohama-cu.ac.jp }\\
  ${}^{*,\dagger,\ddagger,\P}$Department of Mathematical Sciences\\
  Yokohama City University, 
  Yokohama, 236--0027, 
  Japan\\
  ${}^\S$Department of Mathematical Sciences\\
  Waseda University, 
  Tokyo, 169--8555, 
  Japan\\
  }
\date{}
\maketitle
%
%
%
%
\begin{abstract}
  In this letter the explicit form of evolution operator for the four atoms 
  Tavis--Cummings model is given. 
\end{abstract}
%


%
%
%
%


This paper is a sequel to the papers \cite{papers-2} and the purpose is to 
give an explicit form to the evolution operator of Tavis--Cummings model 
(\cite{TC}) with four atoms. 
This model is a very important one in Quantum Optics and has been studied 
widely, see \cite{books} as general textbooks in quantum optics. 

We are studying a quantum computation and therefore want to study the model 
from this point of view, namely the quantum computation based on atoms of 
laser--cooled and trapped linearly in a cavity. We must in this model 
construct a controlled NOT gate or other controlled unitary gates to perform 
a quantum computation, see \cite{KF1} as a general introduction to 
this subject. 

For that aim we need the explicit form of evolution operator of the models 
with one, two and three atoms (at least). As to the model of one atom or 
two atoms it is more or less known (see \cite{papers-1}), 
while it was recently given by us \cite{papers-1} for the three atoms. 

Since we moreover succeeded in finding the explicit form for the case of 
four atoms we report it. 

\vspace{5mm}
The Tavis--Cummings model (with $n$--atoms) that we will treat in this paper 
can be written as follows (we set $\hbar=1$ for simplicity). 
\begin{equation}
\label{eq:hamiltonian}
H=
\omega {1}_{L}\otimes a^{\dagger}a + 
\frac{\Delta}{2} \sum_{i=1}^{n}\sigma^{(3)}_{i}\otimes {\bf 1} +
g\sum_{i=1}^{n}\left(
\sigma^{(+)}_{i}\otimes a+\sigma^{(-)}_{i}\otimes a^{\dagger} \right),
\end{equation}
where $\omega$ is the frequency of radiation field, $\Delta$ the energy 
difference of two level atoms, $a$ and $a^{\dagger}$ are 
annihilation and creation operators of the field, and $g$ a coupling constant, 
and $L=2^{n}$. Here $\sigma^{(+)}_{i}$, $\sigma^{(-)}_{i}$ and 
$\sigma^{(3)}_{i}$ are given as 
\begin{equation}
\sigma^{(s)}_{i}=
1_{2}\otimes \cdots \otimes 1_{2}\otimes \sigma_{s}\otimes 1_{2}\otimes \cdots 
\otimes 1_{2}\ (i-\mbox{position})\ \in \ M(L,\fukuso)
\end{equation}
where $s$ is $+$, $-$ and $3$ respectively and 
\begin{equation}
\label{eq:sigmas}
\sigma_{+}=
\left(
  \begin{array}{cc}
    0& 1 \\
    0& 0
  \end{array}
\right), \quad 
\sigma_{-}=
\left(
  \begin{array}{cc}
    0& 0 \\
    1& 0
  \end{array}
\right), \quad 
\sigma_{3}=
\left(
  \begin{array}{cc}
    1& 0  \\
    0& -1
  \end{array}
\right), \quad 
1_{2}=
\left(
  \begin{array}{cc}
    1& 0  \\
    0& 1
  \end{array}
\right).
\end{equation}

Here let us rewrite the hamiltonian (\ref{eq:hamiltonian}). If we set 
\begin{equation}
\label{eq:large-s}
S_{+}=\sum_{i=1}^{n}\sigma^{(+)}_{i},\quad 
S_{-}=\sum_{i=1}^{n}\sigma^{(-)}_{i},\quad 
S_{3}=\frac{1}{2}\sum_{i=1}^{n}\sigma^{(3)}_{i},
\end{equation}
then (\ref{eq:hamiltonian}) can be written as 
\begin{equation}
\label{eq:hamiltonian-2}
H=
\omega {1}_{L}\otimes a^{\dagger}a + \Delta S_{3}\otimes {\bf 1} + 
g\left(S_{+}\otimes a + S_{-}\otimes a^{\dagger} \right)
\equiv H_{0}+V,
\end{equation}
which is very clear. We note that $\{S_{+},S_{-},S_{3}\}$ satisfy the 
$su(2)$--relation 
\begin{equation}
[S_{3},S_{+}]=S_{+},\quad [S_{3},S_{-}]=-S_{-},\quad [S_{+},S_{-}]=2S_{3}.
\end{equation}
However, the representation $\rho$ defined by 
$
\rho(\sigma_{+})=S_{+},\ \rho(\sigma_{-})=S_{-},\ 
\rho(\sigma_{3}/2)=S_{3}
$
is a reducible representation of $su(2)$. 

We would like to solve the Schr{\" o}dinger equation 
\begin{equation}
\label{eq:schrodinger}
i\frac{d}{dt}U=HU=\left(H_{0}+V\right)U, 
\end{equation}
where $U$ is a unitary operator (called the evolution operator). 
We can solve this equation by using the {\bf method of constant variation}. 
The result is well--known to be 
\begin{equation}
\label{eq:full-solution}
U(t)=\left(\mbox{e}^{-it\omega S_{3}}\otimes 
\mbox{e}^{-it\omega a^{\dagger}a}\right)
\mbox{e}^{-itg\left(S_{+}\otimes a + S_{-}\otimes a^{\dagger}\right)}
\end{equation}
under the resonance condition $\Delta=\omega$, 
where we have dropped the constant unitary operator for simplicity. 
Therefore 
we have only to calculate the term (\ref{eq:full-solution}) explicitly, 
which is however a very hard task \footnote{the situation is very similar to 
that of the paper quant-ph/0312060 in \cite{qudit-papers}}. 
In the following we set 
\begin{equation}
\label{eq:A}
A_{n}=S_{+}\otimes a + S_{-}\otimes a^{\dagger}
\end{equation}
for simplicity. 
We can determine\ $\mbox{e}^{-itgA_{n}}$\ for $n=4$ (four atoms case) 
completely. As to the cases of $n=1\sim 3$ see \cite{papers-2}.

\vspace{3mm}
\par \noindent 
{\bf Four Atoms Case}\quad In this case $A_{4}$ in (\ref{eq:A}) is written as 
\begin{equation}
\label{eq:A-four}
A_{4}=
\left(
  \begin{array}{cccccccccccccccc}
    0 & a & a & 0 & a & 0 & 0 & 0 & a &   &   &   &   &   &   &   \\
    a^{\dagger}& 0 & 0 & a & 0 & a & 0 & 0 &   & a &   &   &   &   &   &  \\
    a^{\dagger}& 0 & 0 & a & 0 & 0 & a & 0 &   &   & a &   &   &   &   &  \\
    0 & a^{\dagger}& a^{\dagger} & 0 & 0 & 0 & 0 & a &   &   &   & a &  &  
    &   &    \\
    a^{\dagger}& 0 & 0 & 0 & 0 & a & a & 0 &   &   &   &   & a &   &  &   \\
    0 & a^{\dagger}& 0 & 0 & a^{\dagger} & 0 & 0 & a &  &  &  &  &  & a &  
    &  \\
    0 & 0 & a^{\dagger} & 0  & a^{\dagger} & 0 & 0 & a &  &  &  &  &  &  & 
    a &   \\
    0 & 0 & 0 & a^{\dagger} & 0 & a^{\dagger} & a^{\dagger} & 0 &  &  &  &  & 
    &  &  & a  \\
    a^{\dagger}&  &  &  &  &  &  &  & 0 & a & a & 0 & a & 0 & 0 & 0  \\
      & a^{\dagger}&  &  &  &  &  &  & a^{\dagger}& 0 & 0 & a & 0 & a & 0 & 0 
    \\
      &  & a^{\dagger}&  &  &  &  &  & a^{\dagger}& 0 & 0 & a & 0 & 0 & a & 0 
    \\
      &  &  & a^{\dagger}&  &  &  &  & 0 & a^{\dagger}& a^{\dagger} & 0 & 0 & 
    0 & 0 & a \\
      &  &  &  & a^{\dagger}&  &  &  & a^{\dagger}& 0 & 0 & 0 & 0 & a & a & 0 
    \\
      &  &  &  &  & a^{\dagger}&  &  & 0 & a^{\dagger}& 0 & 0 & a^{\dagger} & 
    0 & 0 & a \\
      &  &  &  &  &  & a^{\dagger}&  & 0 & 0 & a^{\dagger} & 0  & a^{\dagger} 
    & 0 & 0 & a \\
      &  &  &  &  &  &  & a^{\dagger} & 0 & 0 & 0 & a^{\dagger} & 0 & 
    a^{\dagger} & a^{\dagger} & 0 
  \end{array}
\right).
\end{equation}

If we set $T$ as 
\begin{eqnarray*}
&&T= \\
&&\hspace{-5mm}\\
&&
\left(
\begin{array}{cccccccccccccccc}
0 & 0 & 0 & 0 & 0 & 0 & 0 & 0 & 0 & 0 & 0 & 1 & 0 & 0 & 0 & 0 \\
0 & \frac{1}{\sqrt{2}} & 0 & 0 & 0 & \frac{1}{\sqrt{6}} & 0 & 0 & 
\frac{1}{2\sqrt{3}} & 0 & 0 & 0 & \frac{1}{2} & 0 & 0 & 0 \\
0 & -\frac{1}{\sqrt{2}} & 0 & 0 & 0 & \frac{1}{\sqrt{6}} & 0 & 0 & 
\frac{1}{2\sqrt{3}} & 0 & 0 & 0 & \frac{1}{2} & 0 & 0 & 0 \\
0 & 0 & 0 & 0 & \frac{1}{\sqrt{3}} & 0 & \frac{1}{\sqrt{3}} & 0 & 0 & 
\frac{1}{\sqrt{6}} & 0 & 0 & 0 & \frac{1}{\sqrt{6}} & 0 & 0 \\
0 & 0 & 0 & 0 & 0 & -\sqrt{\frac{2}{3}} & 0 & 0 & \frac{1}{2\sqrt{3}} & 0 & 
0 & 0 & \frac{1}{2} & 0 & 0 & 0 \\
\frac{1}{2} & 0 & \frac{1}{2} & 0 & -\frac{1}{2\sqrt{3}} & 0 & 
-\frac{1}{2\sqrt{3}} & 0 & 0 & \frac{1}{\sqrt{6}} & 0 & 0 & 0 & 
\frac{1}{\sqrt{6}} & 0 & 0 \\
-\frac{1}{2} & 0 & -\frac{1}{2} & 0 & -\frac{1}{2\sqrt{3}} & 
0 & -\frac{1}{2\sqrt{3}} & 0 & 0 & \frac{1}{\sqrt{6}} & 0 & 0 & 
0 & \frac{1}{\sqrt{6}} & 0 & 0 \\
0 & 0 &  0 & 0 & 0 & 0 & 0 & 0 & 0 & 0 & \frac{\sqrt{3}}{2} & 
0 & 0 & 0 & \frac{1}{2} & 0 \\
0 & 0 & 0 & 0 & 0 & 0 & 0 & 0 & -\frac{\sqrt{3}}{2} & 0 & 
0 & 0 & \frac{1}{2} & 0 & 0 & 0 \\
-\frac{1}{2} & 0 & \frac{1}{2} & 0 & -\frac{1}{2\sqrt{3}} & 0 & 
\frac{1}{2\sqrt{3}} & 0 & 
0 & -\frac{1}{\sqrt{6}} & 0 & 0 & 0 & \frac{1}{\sqrt{6}} & 0 & 0 \\
\frac{1}{2} & 0 & -\frac{1}{2} & 0 & -\frac{1}{2\sqrt{3}} & 0 & 
\frac{1}{2\sqrt{3}} & 0 & 0 & -\frac{1}{\sqrt{6}} & 0 & 0 & 
0 & \frac{1}{\sqrt{6}} & 0 & 0 \\
0 & 0 & 0 & 0 & 0 & 0 & 0 & \sqrt{\frac{2}{3}} & 0 & 0 & 
-\frac{1}{2\sqrt{3}} & 0 & 0 & 0 & \frac{1}{2} & 0 \\
0 & 0 & 0 & 0 & \frac{1}{\sqrt{3}} & 0 & -\frac{1}{\sqrt{3}} & 
0 & 0 & -\frac{1}{\sqrt{6}} & 
0 & 0 & 0 & \frac{1}{\sqrt{6}} & 0 & 0 \\
0 & 0 & 0 & \frac{1}{\sqrt{2}} & 0 & 0 & 0 & -\frac{1}{\sqrt{6}} & 
0 & 0 & -\frac{1}{2\sqrt{3}} & 
0 & 0 & 0 & \frac{1}{2} & 0 \\
0 & 0 & 0 & -\frac{1}{\sqrt{2}} & 0 & 0 & 0 & -\frac{1}{\sqrt{6}} & 
0 & 0 & -\frac{1}{2\sqrt{3}} & 0 & 0 & 0 & \frac{1}{2} & 0 \\
0 & 0 & 0 & 0 & 0 & 0 & 0 & 0 & 0 & 0 & 0 & 0 & 0 & 0 & 0 & 1
\end{array}
\right) 
\end{eqnarray*}
then it is not difficult to see 
\begin{eqnarray*}
&& T^{\dagger} A_4 T=\\
&& \hspace{-10mm}
\left(
\begin{array}{cccccccccccccccc}
	0&&&&&&&&&&&&&&& \\
	&0&{\sqrt{2}} a&0&&&&&&&&&&&& \\
	&{\sqrt{2}} a^{\dagger}&0&{\sqrt{2}} a&&&&&&&&&&&& \\
	&0&{\sqrt{2}} a^{\dagger}&0&&&&&&&&&&&& \\
	&&&&0&&&&&&&&&&& \\
	&&&&&0&{\sqrt{2}} a&0&&&&&&&& \\
	&&&&&{\sqrt{2}} a^{\dagger}&0&{\sqrt{2}} a&&&&&&&& \\
	&&&&&0&{\sqrt{2}} a^{\dagger}&0&&&&&&&& \\
	&&&&&&&&0&{\sqrt{2}} a&0&&&&& \\
	&&&&&&&&{\sqrt{2}} a^{\dagger}&0&{\sqrt{2}} a&&&&& \\
	&&&&&&&&0&{\sqrt{2}} a^{\dagger}&0&&&&& \\
	&&&&&&&&&&&0&2 a&0&0&0 \\
	&&&&&&&&&&&2 a^{\dagger}&0&{\sqrt{6}} a&0&0 \\
	&&&&&&&&&&&0&{\sqrt{6}} a^{\dagger}&0&{\sqrt{6}} a&0 \\
	&&&&&&&&&&&0&0&{\sqrt{6}} a^{\dagger}&0&2 a \\
	&&&&&&&&&&&0&0&0&2 a^{\dagger}&0 
\end{array}
\right)  \\
&&\equiv 0\oplus B_{1}\oplus 0\oplus B_{1}\oplus B_{1}\oplus B_{2}. 
\end{eqnarray*}
This means a well--known decomposition of spin 
\[
\frac{1}{2}\otimes \frac{1}{2}\otimes \frac{1}{2}\otimes \frac{1}{2}
=\left( 0\oplus 1\right) \otimes \frac{1}{2}\otimes \frac{1}{2}
=\left( \frac{1}{2}\oplus \frac{1}{2}\oplus \frac{3}{2}\right) 
\otimes \frac{1}{2}=0\oplus 1\oplus 0\oplus 1\oplus 1\oplus 2. 
\]
Since we have calculated $\mbox{e}^{-itgB_{1}}$ in \cite{papers-2} 
we have only to do $\mbox{e}^{-itgB_{2}}$, which is however very hard. 
The result is 
\begin{eqnarray}
&&\mbox{exp}\left(-itgB_{2} \right)= \nonumber \\
&&{} \nonumber \\
&&
\left(
\begin{array}{ccccc}
f_2(N+2) & 0 & h_1(N+2)a^2 & 0 & k_0(N+2) a^4 \\
0 & f_1(N+1) & 0 & h_0(N+1)a^2 & 0 \\
h_1(N) (a^{\dagger})^2 & 0 & f_0(N) & 0 & h_{-1}(N)a^2 \\
0 & h_0(N-1) (a^{\dagger})^2 & 0 & f_{-1}(N-1) & 0 \\
k_0(N-2) (a^{\dagger})^4 & 0 & h_{-1}(N-2) (a^{\dagger})^2 & 0 & f_{-2}(N-2)
\end{array}
\right)+     \nonumber \\
&&{}         \nonumber \\
&& \hspace{-13mm}
\left(
\begin{array}{ccccc}
 0 & -2iF_1(N+2)a & 0 & -2iH_0(N+2)a^3 & 0 \\
 -2iF_1(N+1) a^{\dagger} & 0 & -\frac{i}{2}H_1(N+1)a & 0 & -2iH_0(N+1)a^3 \\
 0 & -\frac{i}{2}H_1(N) a^{\dagger} & 0 & -\frac{i}{2}H_{-1}(N)a & 0 \\
 -2iH_0(N-1) (a^{\dagger})^3 & 0 & -\frac{i}{2}H_{-1}(N-1) a^{\dagger} & 0 & 
 -2iF_{-1}(N-1)a \\
 0 & -2iH_0(N-2) (a^{\dagger})^3 & 0 & -2iF_{-1}(N-2) a^{\dagger} & 0
\end{array}
\right)     \nonumber \\
&& 
\end{eqnarray}
where 
\begin{eqnarray*}
&&f_2 \equiv 
1+4(N-1)\{ (u_+/\lambda_+) (\cos tg\sqrt{\lambda_+}-1)
 -(u_-/\lambda_-) (\cos tg\sqrt{\lambda_-}-1) \} 
 /\sqrt{d} \\
&&f_1 \equiv 
(u_{+} \cos tg\sqrt{\lambda_{+}}-u_{-} \cos tg\sqrt{\lambda_{-}})
 /\sqrt{d}, \\
&&f_0 \equiv 
1+2\{ (v_{+}w_{+}/\lambda_{+}) (\cos tg\sqrt{\lambda_{+}}-1)
 -(v_{-}w_{-}/\lambda_{-}) (\cos tg\sqrt{\lambda_{-}}-1) \}
 /\sqrt{d}, \\
&&f_{-1} \equiv 
 (u_{+} \cos tg\sqrt{\lambda_{-}}
 -u_{-} \cos tg\sqrt{\lambda_{+}})/\sqrt{d}, \\
&&f_{-2} \equiv 
1+4(N+2) \{ (u_{+}/\lambda_{-}) (\cos tg\sqrt{\lambda_{-}}-1)
 -(u_{-}/\lambda_{+}) (\cos tg\sqrt{\lambda_{+}}-1) \} /\sqrt{d}, \\
&&h_1 \equiv 
 2\{ (v_{+}/\lambda_{+}) (\cos tg\sqrt{\lambda_{+}}-1)
  -(v_{-}/\lambda_{-}) (\cos tg\sqrt{\lambda_{-}}-1) \} /\sqrt{d}, \\
&&h_0 \equiv 
 (\cos tg\sqrt{\lambda_{+}}-\cos tg\sqrt{\lambda_{-}})/\sqrt{d}, \\
&&h_{-1} \equiv 
 2 \{ (w_{+}/\lambda_{+}) (\cos tg\sqrt{\lambda_{+}}-1)
 -(w_{-}/\lambda_{-}) (\cos tg\sqrt{\lambda_{-}}-1) \} /\sqrt{d}, \\
&&k_0 \equiv 
 4 \{ (1/\lambda_{+}) (\cos tg\sqrt{\lambda_{+}}-1)
 -(1/\lambda_{-}) (\cos tg\sqrt{\lambda_{-}}-1) \} /\sqrt{d}, 
\end{eqnarray*}
and
\begin{eqnarray*}
&&F_1 \equiv 
\{(u_{+}/\sqrt{\lambda_{+}}) \sin tg\sqrt{\lambda_{+}}-
  (u_{-}/\sqrt{\lambda_{-}}) \sin tg\sqrt{\lambda_{-}}\}/\sqrt{d}, \\
&&F_{-1} \equiv 
\{(u_{+}/\sqrt{\lambda_{+}}) \sin tg\sqrt{\lambda_{-}}-
  (u_{-}/\sqrt{\lambda_{-}}) \sin tg\sqrt{\lambda_{+}}\}/\sqrt{d}, \\
&&H_1 \equiv 
2\{(v_{+}/\sqrt{\lambda_{+}}) \sin tg\sqrt{\lambda_{+}}-
   (v_{-}/\sqrt{\lambda_{-}}) \sin tg\sqrt{\lambda_{-}}\}/\sqrt{d}, \\
&&H_0 \equiv 
\{(1/\sqrt{\lambda_{+}}) \sin tg\sqrt{\lambda_{+}}-
  (1/\sqrt{\lambda_{-}}) \sin tg\sqrt{\lambda_{-}}\}/\sqrt{d}, \\
&&H_{-1} \equiv 
2\{(w_{+}/\sqrt{\lambda_{+}}) \sin tg\sqrt{\lambda_{+}}-
   (w_{-}/\sqrt{\lambda_{-}}) \sin tg\sqrt{\lambda_{-}}\}/\sqrt{d}
\end{eqnarray*}
, and $d=d(N)$, $\lambda_{\pm}=\lambda_{\pm}(N)$, $u_{\pm}=u_{\pm}(N)$, 
$v_{\pm}=v_{\pm}(N)$ and $w_{\pm}=w_{\pm}(N)$ defined by 
\begin{eqnarray*}
&&\lambda_{\pm}(N) \equiv 10N+5 \pm 3\sqrt{d(N)},\ 
u_{\pm}(N) \equiv \frac 12 (-3 \pm \sqrt{d(N)}),               \\
&&v_{\pm}(N) \equiv \sqrt{\frac 32}(2N-1 \pm \sqrt{d(N)}),\ 
w_{\pm}(N) \equiv \sqrt{\frac 32}(2N+3 \pm \sqrt{d(N)}),       \\
&&d(N) \equiv 4N^2+4N+9. 
\end{eqnarray*}

This form is very complicated. We note that in the process of calculation 
we used Mathematica to the fullest. 

\vspace{5mm}
A comment is in order. We would like to generalize the results in this paper 
and \cite{papers-2} to the cases of more than four atoms. However, it is not 
easy to perform it due to some technical reasons. There is a (big ?) gap 
between the four atoms and the five ones.

\vspace{10mm}
We obtained the explicit form of evolution operator of the Tavis--Cummings 
model for the case of three and four atoms, so many applications to 
quantum optics or mathematical physics will be expected, see for example 
\cite{papers-1}. 
In the near future we will apply the result to a quantum computation based on 
atoms of laser--cooled and trapped linearly in a cavity \cite{FHKW}. 

We conclude this paper by making a comment. The Tavis--Cummings model 
is based on (only) two energy levels of atoms. However, an atom has in general 
infinitely many energy levels, so it is natural to use this possibility. 
We are also studying a quantum computation based on multi--level systems of 
atoms (a qudit theory) \cite{qudit-papers}. Therefore we would like to extend 
the Tavis--Cummings model based on two--levels to a model based on 
multi--levels. This is a very challenging task.


\end{document}